\documentclass[sigconf]{acmart}
    \usepackage[utf8]{inputenc}
    \usepackage{tikz}
    \usepackage[bold]{hhtensor}

    \usepackage{verbatimbox}

\copyrightyear{2020} 
\acmYear{2020} 
\setcopyright{acmlicensed}
\acmConference[SIGCSE '20]{The 51st ACM Technical Symposium on Computer Science Education}{March 11--14, 2020}{Portland, OR, USA}
\acmBooktitle{The 51st ACM Technical Symposium on Computer Science Education (SIGCSE '20), March 11--14, 2020, Portland, OR, USA}
\acmPrice{15.00}
\acmDOI{10.1145/3328778.3366952}
\acmISBN{978-1-4503-6793-6/20/03}

\begin{document}

\fancyhead{}
\title{Teaching Quantum Computing through a Practical Software-driven Approach}
\subtitle{Experience Report}

\author{Mariia Mykhailova}
\email{mamykhai@microsoft.com}
\affiliation{Microsoft Quantum, Microsoft, Redmond, WA, USA}

\author{Krysta M.~Svore}
\email{ksvore@microsoft.com}
\affiliation{Microsoft Quantum, Microsoft, Redmond, WA, USA}


\begin{abstract}
Quantum computing harnesses quantum laws of nature to enable new types of algorithms, not efficiently possible on traditional computers, that may lead to breakthroughs 
in crucial areas like materials science and chemistry. There is rapidly growing demand for a quantum workforce educated in the basics of quantum computing, 
in particular in quantum programming. However, there are few offerings for non-specialists and little information on best practices for training 
 computer science and engineering students.

In this report we describe our experience teaching an undergraduate course on quantum computing using a practical, software-driven approach. 
We centered our course around teaching quantum algorithms through hands-on programming, reducing the significance of traditional written assignments 
and relying instead on self-paced programming exercises (``Quantum Katas''), a variety of programming assignments, and a final project. 
We observed that the programming sections of the course helped students internalize theoretical material presented during the lectures. 
In the survey results, students indicated that the programming exercises and the final project contributed the most to their learning process. 

We describe the motivation for centering the course around quantum programming, discuss major artifacts used in this course, 
and present our lessons learned and best practices for a future improved course offering. 
We hope that our experience will help guide instructors who want to adopt a practical approach 
to teaching quantum computing and will enable more undergraduate programs to offer quantum programming as an elective.
\end{abstract}

\maketitle

\section{Introduction}

Quantum computing is an emerging computing paradigm which relies on quantum-mechanical phenomena such as superposition and entanglement to perform computations.
It enables new types of algorithms that may solve computational problems that are intractable on traditional digital computers.
Quantum computing is expected to lead to breakthroughs in crucial areas like machine learning and optimization, materials science, and chemistry \cite{nwchem}. 

At present, quantum computing is just emerging from its infancy.  Similar to the early days of ``classical'' computing, 
scientific visionaries created a model of computation and supporting theories on the kinds of problems that may be efficiently solved using the new quantum paradigm; 
and companies such as Microsoft, IBM, and Google invest in developing hardware to bring this vision to life.  

Even at this stage, there is already an appetite for a diverse workforce trained in quantum computing - not just theoretical algorithms designers, 
but also software and hardware engineers who can create quantum software development tools, integrate them with the hardware, and develop practical algorithms for specific applications.  
Organizations are starting to realize the benefits of hiring in-house experts to adapt quantum algorithms for solving domain-specific problems, 
to identify the impact of quantum computing advances on their domain, and to integrate the learnings into improved business solutions. 
Quantum computing courses focused on programming and real-world applications are necessary to foster such a generation of quantum software engineers.

For the computer science education system, quantum computing poses a different challenge: what is the best way to train the future quantum workforce?
Very few universities offer training in quantum computing, and for those that do offer it, most curriculum to date is focused on graduate students. 
A lot of the courses cover the details of the underlying quantum physics and the hardware, which, while fascinating in their own right, are not required for learning quantum computing, 
just as a course on integrated chip design is not required to study classical algorithms. 
Finally, the traditional approach to teaching quantum computing focuses on purely theoretical aspects of the field, drawing very little content from practical applications and programming quantum computers. 

In this report we describe our experience teaching an undergraduate course 
``CSE490Q: Introduction to Quantum Computing and Quantum Programming in Q\#'' 
at the University of Washington \cite{uw}. We aimed to make this course practical and accessible to a broader academic audience.
We outline the guiding principles behind our course design, describe the major artifacts created for the course, and discuss the lessons learned.  
We hope that our experience will inspire more instructors to offer quantum computing courses following a similar software-driven approach. 
We also hope that our work enables a broad set of instructors and students to engage in quantum programming, for lack of this engagement will lead to a growing workforce shortage.

\section{Background}

A traditional approach to teaching quantum computing focuses on theoretical presentation of the field.
Quantum computing is commonly taught from within a physics department, with an emphasis on the theoretical physics of quantum systems and hardware devices 
\cite{caltech, berkeley-phys}.
Alternatively, a lot of courses focus on the mathematical aspects of information theory \cite{berkeley-math, waterloo}.
In both described scenarios the students' performance is evaluated based on written assignments and exams and final projects in the form of an oral presentation 
and a written report on the selected algorithm.

Majority of the aforementioned courses target graduate students; 
while undergraduate quantum computing courses taught within computer science departments exist \cite{ut-austin, ucla}, they remain rare outliers. 

Finally, only in the last years quantum programming has found its way into some of the introductory quantum computing courses \cite{ucla}; 
majority of the courses still do not seek to train students in the practical aspects of quantum computing.

In contrast, computer science departments teach classical computing by introducing a programming language and challenging students to solve simple problems with it, 
moving on to classical data structures and algorithms and implementing them in that programming language, 
and then learning advanced programming and topics of special interest by offering advanced electives.  
It is never taught by discussing the physics of a transistor and computational complexity theory without ever doing any hands-on programming.

\section{Motivation}

We aimed to create a course that started with programming and emphasized hands-on learning to make quantum computing accessible to undergraduate students.  
We defined several principles in the design and delivery of our course:

\begin{itemize}
   \item 
\textbf{Teach quantum computing through software engineering instead of physics.}
    We approached the course as computer scientists, deliberately avoiding physics in the lectures. We presented the qubits as abstract objects described by vectors instead of quantum mechanical states, and 
    the operations on them as matrix transformations instead of physical processes. We dedicated one lecture at the end to an overview of possible physical implementations of quantum hardware, 
    since the hardware is both emerging and intriguing, but otherwise we kept the discussion to mathematical abstractions and concepts and their programming counterparts.

    \item 
\textbf{Teach quantum algorithms through programming.} 
    Traditional computer science courses teach algorithms in a practical way - by requiring that students implement them using one of the existing programming languages.
    While quantum computing is not yet at the stage in which algorithms of practical interest can be executed on real quantum hardware, 
    we developed programming assignments that can be implemented and executed on a quantum simulator 
    that runs on classical computers and simulates the programmed behavior of a quantum system. 
    Such assignments allow to learn to implement and debug quantum algorithms and impart deeper understanding of quantum algorithm design. 

    \item 
\textbf{Focus on practical aspects of algorithms, not mathematical proofs.}
    Rigorous proofs are a traditional part of an in-depth quantum computing course. However, we feel that they are not necessary for an introductory software-oriented course targeted at 
    students seeking introductory knowledge about quantum computing.
    Reducing the mathematical rigor and replacing it with quantum programming and software design allows to focus on the practical aspects of the algorithms' implementation, 
    as well as to reduce the amount of prerequisite algebra for the course, making it more accessible to undergraduate students.
\end{itemize}

Since we designed our course with a heavy emphasis on quantum algorithms implementation, we had to identify appropriate tools for the hands-on learning part of the course.
We chose to use a single programming language and environment throughout the course.  
We taught the course using Q\#\cite{rwdsl2018}, a high-level domain-specific programming language used for expressing quantum algorithms, and the Microsoft Quantum Development Kit (QDK) \cite{docs}, 
an open-source software development kit that includes a compiler, a general purpose quantum simulator, and other development tools essential for quantum software design and development, 
such as specialized quantum simulators, debugger and algorithm resource estimator.

\section{Curriculum}

The course covered a combination of introductory topics featured in most ``Introduction to quantum computing'' courses 
\cite{caltech, ut-austin} or textbooks \cite{mermin, nielsen-chuang, yanofsky},
quantum routines used as building blocks for more complicated algorithms, and quantum computing applications.  The core topics included:

\begin{itemize}
    \item Fundamental concepts of quantum computing: qubits, quantum gates, superposition, entanglement, measurements.
    \item Simple algorithms: teleportation, Deutsch and Deutsch–Jozsa algorithms, Simon's algorithm.
    \item Quantum oracles.
    \item Grover's search algorithm.
    \item Quantum routines: Fourier transform, amplitude amplification, phase estimation.
    \item Quantum arithmetic.
    \item Application: order finding and Shor's algorithm for integer factorization.
    \item Application: quantum simulation and quantum chemistry.
    \item Quantum error correction.
\end{itemize}

Most topics involved learning both theoretical properties of the algorithm and its Q\# implementation. 
We made an exception for more advanced applications of quantum computing, focusing mostly on their theoretical aspects, since they are more demanding in terms of prerequisites and more complicated to implement.

The course targeted 3rd and 4th year undergraduate students with previous exposure to linear algebra and programming. 
While most students took the course for a grade, students also had the option to take the course as pass/fail or audit.  

\subsection{Course structure}

The course ran for 10 weeks, with two 80-minutes lectures per week. In addition to the lectures, 
we hosted weekly office hours with the professor and the teaching assistants. 

In the lectures we focused on teaching the fundamental concepts and describing quantum algorithms, and at home learning was focused on learning quantum programming.  
To supplement the lectures, we recommended reading sections of the textbooks \cite{mermin} and \cite{nielsen-chuang}. 
For select lectures, we also identified supplementary reading material - other existing course notes \cite{caltech, berkeley-math} or research papers from xxx.arxiv.org.
While the course had no required programming lab section, we expected students to supplement their learning using a unique open-source project called ``Quantum Katas''
\cite{quantum-katas}.

Student performance was evaluated using the following types of graded assignments:

\begin{itemize}
    \item written assignments (20\% of the final grade)
    \item programming assignments (30\% of the final grade)
    \item final programming-driven project (50\% of the final grade)
\end{itemize}

One of the programming assignments was completion of an online quantum programming competition.
We chose not to hold a mid-term or a final written exam, since we felt that a combination of programming assignments and the final project would better suit the intended course goals.

\subsection{Written assignments}

Written assignments were designed to encourage deeper engagement with the topics covered in the lectures. 
Each assignment consisted of several problems which required calculations and sometimes short proofs.
Each assignment was expected to take around 8 hours for an average student to complete.  
 
The course had 4 written assignments, posted biweekly during the first 8 weeks of the course. 
The students had 2 weeks to complete each assignment and could work in teams.
The written assignments were graded manually by the teaching assistants, based on the reference solution write-ups provided by the course instructors.

\begin{figure}[b]
    \textit{Compute the unitary matrix corresponding to the following \newline quantum circuit:}

    \includegraphics[width=0.4\linewidth]{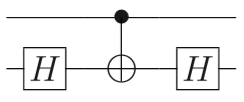}
    \caption{\label{fig-homework-circuits} An example of a written assignment task}
    \Description{An example of a written assignment task: compute the unitary matrix corresponding to a simple quantum circuit}
\end{figure}

An example assignment problem is given in Figure \ref{fig-homework-circuits}.

\subsection{The Quantum Katas}

A unique element of our course was the use of ``Quantum Katas'' \cite{quantum-katas} -
an open-source project aiming to aid learning quantum computing and Q\# programming. 
We created this project and made it available to the broad public prior to start of the course, targeting self-paced learners and quantum computing enthusiasts.
This course was the first to employ the Quantum Katas for a university course in a systematic way. We developed one new kata on phase estimation especially for the course, 
since this is a fairly advanced topic.

The \textit{katas} are sets of programming tasks that cover one topic (e.g., quantum measurements) 
or several related topics (e.g., quantum oracles and Grover's search algorithm).
The Quantum Katas follow several learning principles:
\begin{itemize}
    \item 
\textbf{Active learning.} 
    The tasks of the katas are purely practical - they don't require writing any proofs or drawing any circuits.
    Instead, each task describes a problem and asks the learner to write a fragment of Q\# code to implement the solution.

    \item 
\textbf{Immediate automated feedback.} 
    It is well-known that timely feedback is a critical component of the learning process.
    The primary goal of the Quantum Katas is to automate feedback delivery to learners who might not have easy access to other sources of feedback (such as a teacher).
    
    Each kata includes a testing framework which validates the tasks' solutions as soon as they are written.
    The frameworks use the full-state quantum simulator included in the QDK
    to simulate a quantum program that sets up the inputs required by the task, runs the solution and processes the results.
    This enables the learner to solve the katas on a classical computer without access to quantum hardware.

    Each kata aims to provide as much details about incorrect solutions as the nature of the tasks allows;
    in some topics it is just a ``correct''/``incorrect'' indication, 
    in others it is the percentage of the inputs on which the solution returned an incorrect answer,
    and in some topics the learner gets a full report on the test case on which their solution failed.

    \item 
\textbf{Incremental increase of task complexity.}    
    The tasks in each kata start simple and gradually increase in complexity, building on the previous tasks.
    As the learner completes each task, this success gives them both a confirmation that they have mastered the necessary knowledge 
    and a boost of confidence for attempting the next, more complicated task.
\end{itemize}

\begin{figure}[b]
    \includegraphics[width=\linewidth]{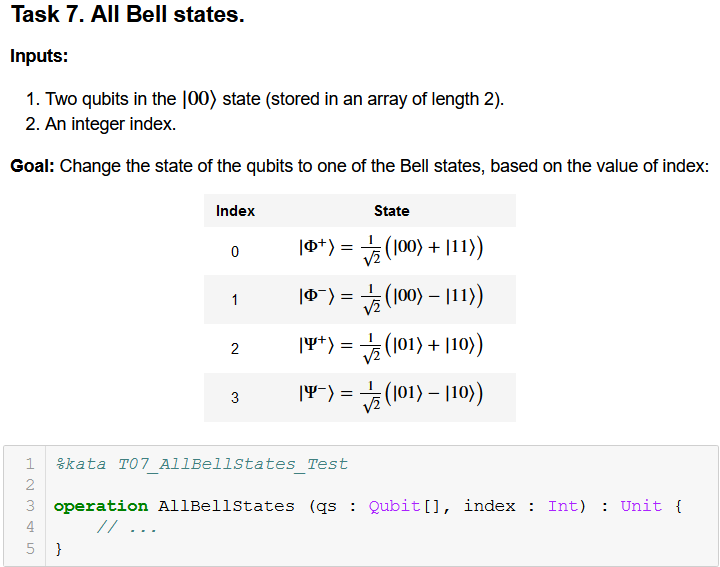}
    \caption{\label{fig-kata-task} ``Bell states'' task from the Superposition kata}
    \Description{``Bell states'' task from the Superposition kata: given two qubits in zero state, change the state of the qubits to one of the Bell states based on the value of an integer parameter}
\end{figure}

Figure \ref{fig-kata-task} shows an example of a task from the Superposition kata - one of the first katas used in the course.  
The goal is to replace $\ldots$ with a Q\# code snippet that solves the problem.

We employed the self-paced nature of the Quantum Katas and their built-in feedback delivery mechanism as a replacement for in-person programming lab time.
Each week we recommended one or several katas to help students practice Q\# programming and internalize the concepts and algorithms introduced that week
necessary to complete the graded assignments.

\subsection{Programming assignments}

The course had 6 programming assignments, posted weekly during the first 6 weeks of the course.
The students had one week to complete each assignment (in one case of a larger assignment the time was extended to two weeks).

\begin{figure}[b]
    \includegraphics[width=\linewidth]{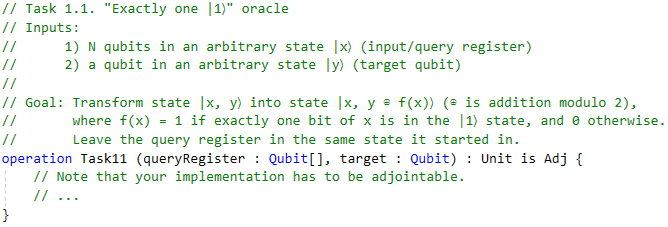}
    \caption{\label{fig-homework-oracles} An example of a programming assignment task}
    \Description{An example of a programming assignment task: given N+1 qubits in an arbitrary state, implement a quantum oracle that corresponds to a classical function "exactly one bit of binary input x is 1"}
\end{figure}

The first 5 assignments were very similar to the Quantum Katas: each one consisted of a set of programming tasks,
with task descriptions and signatures of the operations that needed to be implemented listed in a single boilerplate file. 
The students submitted the Q\# code that would solve the described tasks using the given file structure.
Figure \ref{fig-homework-oracles} shows an example of a task from programming assignment 4, covering oracles for Grover's search algorithm.

These assignments were graded semi-automatically using a framework similar to the one used in the Quantum Katas. 
Each assignment had a testing harness - a Q\# project.
The grading script copied the student's submission into this project, built the project and ran the tests - 
if a test passed, the solution to the corresponding task was considered correct.
The student's score for the assignment was based on the number of tasks that passed the tests.

The sixth programming assignment covered resource estimation - an important practical technique used in quantum algorithms research \cite{nwchem}.
Small instances of quantum algorithms can be simulated on a classical computer to get full information about the system behavior at any point of the execution.
However, larger instances (ones requiring over 50 qubits to run) are impossible to simulate. Resource estimation techniques allow to estimate
the amount of resources (such as the number of qubits, the number of gates and the runtime) necessary to run a large instance of an algorithm.
This is extremely useful both for estimating the parameters of a quantum device that would be required to successfully run a quantum algorithm 
(and thus informing design decisions done by quantum hardware engineers) and for optimizing quantum algorithms.

In the last assignment the students explored the resource estimation tools provided as part of the QDK \cite{trace-simulator}.
They used them to estimate resources, such as number of qubits and number of gates, necessary to run a small instance of Shor's algorithm for integer factorization \cite{shor}.
For this assignment the students submitted a text file with resource counts for a given instance of the algorithm.

\subsection{Programming competition}

The final programming assignment was to participate in the Microsoft Q\# Coding Contest – Winter 2019 \cite{contest} held during the second half of our course.
The contest tasks were prepared by one of the instructors, so we decided to use it as one of the assignments to offer the students a new challenge.

The competition was held online and ran for 72 hours, from Friday morning to Monday morning, and included 13 problems of varying difficulty. 
It covered the following topics:

\begin{description}
    \item [State preparation] The tasks asked to prepare a certain superposition state.
    \item [State discrimination] The tasks asked to distinguish several quantum states.
    \item [Quantum oracles] The tasks asked to develop an oracle that would implement a certain classical function.
    \item [``Unitary patterns''] The tasks had the following form: given a pattern of zero and non-zero elements in a square matrix of size $2^n \times 2^n$, 
    write Q\# code that implements any unitary transformation on $n$ qubits described by a matrix that matches this pattern.
\end{description}

It is worth noting that the first three topics were extensively covered in the Quantum Katas and the programming assignments for the course, though the competition tasks were significantly more challenging.
The last topic was a detour from the usual learning path and required insight into unitary transformations (one of the key concepts in quantum computing).  

To solve each problem, the participants had to write a fragment of Q\# code implementing the required functionality, same as in the Quantum Katas and the programming assignments.
The participants submitted their solutions to the website \cite{contest}, where they would be automatically verified on tests from a predefined set.
The participants received the verification result immediately, so that they could fix incorrect solutions and resubmit them.
The students were graded based only on the number of correctly solved tasks. 
Only one of our students managed to solve all problems, finishing in 22nd place out of 255 participants \cite{contest-results}.

\subsection{Final projects}

Computer science education has an established concept of a capstone project - a multifaceted practical assignment that serves as a culminating academic experience, typically in the end of an academic program.
We selected a capstone project in lieu of a final exam, as it aligned well with our vision of a hands-on, practical introduction to quantum computing.

For the final project, students worked in teams to select and define a problem, develop a solution, write a Q\# program to implement the solution, 
describe their work in a written report, and present it in a 10-minute in-class presentation.
This assignment was designed to teach the students diverse aspects of software development, such as research skills, collaboration, writing and public speaking, in addition to furthering their quantum software development skills.
We had the students self-assemble into teams of three (several students formed two-person teams) and identify a project that they would then work on for 4 weeks.
The project could be a deeper exploration of one of the topics studied in the course, or an independent study of any topic in quantum computing not mentioned in the course.

The project consisted of three graded components:

\textbf{(1) The Q\# code.}    
    In the spirit of the practical approach emphasized throughout the course, the project had to include Q\# code that illustrated or explored the chosen topic. 
    The amount of code had to be about 2-3 programming assignments' worth of effort.

    It is worth noting that the code written for the project has to be able to solve at least some problem instances using the QDK's quantum simulator.
    This added an extra layer of depth to the assignment: the students had to analyze the selected problem to find a small but non-trivial instance 
    that could showcase their implementation.  Some students also chose to demonstrate scaling of their programs using the QDK's resource estimator, 
    which was introduced in the 6th programming assignment. 

    For example, a project that demonstrated solving graph coloring problems using Grover's search algorithm required not only implementing a general solution for the problem, 
    but also using it to solve a small instance.  The team estimated the size of the problem that can be solved using the quantum simulator 
    (around 30 qubits when simulating on a laptop), realized that the number of colors used had to be limited to 4, and then constructed an example graph 
    with non-trivial coloring which was small enough to solve in simulation.  

    This requirement ensured that the students not only practiced Q\# programming and learned new algorithms, but also  
    got a chance to use resource analysis techniques and gained a deeper understanding of the algorithms' properties.

\textbf{(2) The report.}
    The report had to be structured as a short (at most 5 pages) research paper: it had to include an abstract, a problem statement and an overview of existing work, a solution description, 
    and technical results, including implementation details of the code and an analysis of execution results. 
    This requirement was designed to teach scientific communication skills and to further verify the students' understanding of the topic.  

\textbf{(3) The presentation.}
    Two lectures at the end of the course were dedicated to student presentations. Each team had 10 minutes to present a summary of their project and a live demo of the code, 
    with an additional 5 minutes for questions from the other students and course instructors.
    The presentations were a way for students to learn to present their work and communicate the high impact results and 
    for the other students in the class to get an overview of multiple topics not covered in the main part of the course.

The final projects selected by the students spanned a wide range of topics, from simple protocols such as entanglement games and quantum key distribution
and traditional quantum computing problems like solving constraint satisfaction problems using Grover's search algorithm,
to advanced topics like Hamiltonian simulation algorithms for quantum chemistry, 
Bitcoin mining on a quantum computer, quantum approximate optimization algorithm, and quantum binary neural nets.  
Some students implemented existing algorithms described in published papers.  Others developed entirely new algorithms based on their learnings from class.

Many students chose to frame their final projects as a Quantum Kata and even contributed them back to the public open-source repository after the end of the course
\cite{kata-contributions}.
We consider this a major success indicator of our approach - the students began the course with zero knowledge about quantum computing and 
by the end of the course were able to implement quantum algorithms and successfully contribute them back to the project which is used by thousands of other learners worldwide!

\section{Lessons Learned}

We base our learnings on our observations during the course, the students' performance evaluation, and the student feedback collected at the end of the course 
(14 out of 35 students who completed the course submitted the feedback).

It is worth noting that four of our students went on to do internships at tech companies working in quantum computing domain, 
and several students identified that they were interested in pursing the topic for graduate school.

Overall, the majority of the students indicated that the class was very intellectually stimulating, 
in some cases noting that it was the coolest topic they had learned in their studies. 

The students expressed that they liked learning the algorithms through programming. 
They enjoyed solving the Quantum Katas and found that this contributed a lot to the learning process.  
They liked the self-paced nature of the Katas and even replicated the model in some of the final projects.  
Students also appreciated the final projects and found them an excellent way to dive deeper into areas of quantum computing that overlapped with their interests.

While the students preferred the programming assignments to the written assignments, 
the programming tasks got progressively more challenging throughout the course.  
Students shared that the assignments were at times too challenging to complete given the workload they also had from other courses.  
The programming competition was also too hard for most students, since it ran over the weekend, ran for a much shorter period of time compared to regular programming assignments, 
and featured harder problems. The timed nature of the competition also added undue pressure.

Our teaching assistants expressed their appreciation of the automated grading process for programming assignments; anybody who have ever taught a course
knows that grading students' assignments is a significant share of the workload, and automating this process frees the instructors and TAs 
to focus on other parts of the teaching process.

The students worked together often on the written assignments, and in the early weeks the scores were high.  
As the topics became more complex, the mathematical background of the average computer science major proved insufficient to master the lecture material and to complete some of the assignments. 

In general the amount of material covered in this first offering of an undergraduate-level course was too ambitious 
and the workload was too large for the average CS student.  
At the same time the students agreed that the course should be offered again, as the subject is fascinating and of wide interest to CS students today. 
To make the future runs of the course and its students more successful, we suggest some improvements to the course logistics and delivery.

\textbf{Recommend linear algebra and programming as the prerequisites.}    
    The course was supposed to list both linear algebra and programming as the prerequisites, but due to a paperwork error it listed no prerequisites.
    At the beginning of the term we had over 50 students signed up for the class, but after the 1st week almost 20 dropped out of the course.
    Having the right prerequisites helps set students' expectations and allows the instructors focus on the new content, specific to quantum computing.
    Alternatively, the instructors could adjust the course material to spend the first week reviewing the necessary topics in linear algebra.
    
\textbf{Allocate dedicated programming lab time.}   
    The course did not have a dedicated programming lab section, so we relied heavily on students' self-learning, assisted by the language documentation \cite{docs},   
    the existing samples code base \cite{samples} and the Quantum Katas.
    We also had an online channel for reaching out to a Q\# expert (one of the instructors) with language questions, and office hours with both the professor and the TAs. 
    In a future offering of this course, we highly recommend a required weekly 90 or 120 minute programming section.  
    While students self-organized to work on programming assignments together, scheduled lab time with in-person access to an instructor with quantum computing expertise 
    would have greatly improved the students' learning experience.
    
\textbf{Recruit \enspace teaching \enspace assistants \enspace with \enspace quantum computing \newline knowledge.} 
    We hope that eventually this course will help create a pool of students who can assist in teaching the course;
    until then the instructors will have extra responsibilities in helping students with the programming and written assignments.
    
\textbf{Adjust the topics to match the students' level.} 
    Our curriculum covered many topics in quantum computing, more than was necessary for an introduction to the topic.
    For a 10-week undergraduate course, we recommend removing up to three topics to enable deeper understanding of core quantum algorithms.  
    We found that students struggled with the more advanced topics, such as quantum simulation and quantum chemistry, 
    since they require domain-specific knowledge and advanced mathematics.
    
Spending more lecture time on the core concepts of quantum computing, such as the ones introduced in Grover's search algorithm, 
and on Q\# programming will help students master the fundamentals of quantum programming better. 
Advanced applications of quantum computing may be introduced through a brief overview without an accompanying assignment. 
Students with sufficient math background can explore those topics in their final projects, which we observed in multiple cases.

\textbf{Develop a testing harness for the programming assignments to share with the students.}
    We developed a testing harness to grade the students' assignments during the course, which was similar to the one used in the Quantum Katas, i.e., 
    it was a Q\# project which included correct answers to all tasks.  Since the harness revealed the solution, we were not able to distribute it to the students together with the assignment. 

    We observed that access to such a harness would have enabled the students to focus on learning the new content, especially during the first weeks of the course.
    We shared simplified examples of testing harnesses with the students at the beginning of the course to help them learn to write their own tests, 
    but the additional need to come up with a way to verify solutions nearly doubled the amount of work required to complete the assignments.

    Developing a method to test the solutions to programming tasks without revealing the correct answer is a challenge, 
    however, we feel that the students' experience with testing their programming assignments can be improved.
    For a future course, we plan to devise a method to let students test their solutions before submitting them.

\section{Conclusions and future work}

From our experience teaching this course we conclude that our software-oriented approach to teaching quantum computing is a viable way to introduce students to 
the practical aspects of quantum computing. By the end of the course the students develop a firm grasp on the foundations 
of quantum computing and are able to program quantum algorithms for real problems. 
Such an introduction is a perfect fit for an undergraduate course, 
and can be followed by in-depth study of more advanced topics if desired.

We are working on improving the course based on our learnings, and releasing the artifacts created for it to university professors worldwide.
Several instructors in different countries have already used our Quantum Katas and programming assignments in their undergraduate courses
and reported findings similar to ours - the students invariably enjoy the hands-on portion of the course and learn a lot from it.
Enhancing the interaction with the programming assignments, addressing more programming concepts in class, and reducing the content and assignments 
will enable an improved student experience and deeper learnings. We made progress toward our goal of inspiring a generation of quantum programmers, 
launching four students into quantum computing internships and inspiring several others to pursue graduate work in quantum computing.

\begin{acks}

We would like to thank the guest lecturers who helped us prepare and deliver the course: Martin Roetteler, Robin Kothari, Nicolas Delfosse, Michael Beverland and Guang Hao Low.
We would also like to thank our teaching assistants, Logan Weber and Philip Garrison, for their help.
We thank the Computer Science and Engineering department at the University of Washington for enabling this course.
We also thank Linda Lauw.
Finally, a special thanks to the students whose experience and feedback made this course and report possible.

\end{acks}

\bibliographystyle{ACM-Reference-Format}
\bibliography{biblio}

\end{document}